\documentclass{icrc2009}

\usepackage{graphicx}   
\usepackage[caption=false]{caption}    
\usepackage[font=footnotesize]{subfig} 
\usepackage{fixltx2e}
\usepackage{url}

\newcommand{\shorttitle}[1]%
{\markboth{Proceedings of the 31\MakeLowercase{$^{st}$} ICRC, {\L}\'{o}d\'{z} 2009}{#1} }
\newcommand{\etal}{\MakeLowercase{\textit{et al. }}} 

\begin{document}
\title{Simulation study of GZK photon fluxes}

\author{\IEEEauthorblockN{Daniel Kuempel\IEEEauthorrefmark{1},
			  Karl-Heinz Kampert\IEEEauthorrefmark{1} and
                          Markus Risse\IEEEauthorrefmark{1}}
                          \\
\IEEEauthorblockA{\IEEEauthorrefmark{1}Physics Department, University of Wuppertal, Gau\ss str. 20, D-42119 Wuppertal, Germany}}

\shorttitle{D.~Kuempel \etal GZK photon simulation}
\maketitle

\begin{abstract}
The composition of ultra-high energy (UHE) cosmic rays $E>10^{17}$~eV is still unknown. The observation of UHE photons would extend the observed electromagnetic spectrum to highest energy and open a new channel for multimessenger observations in the universe. 
Current limits on the photon flux already constrain ``exotic'' scenarios where a large number of photons is expected by the decay products of supermassive X-particles. Motivated by the growing exposure of UHE cosmic ray experiments - like the Pierre Auger Observatory - the observation of conventionally produced GZK photons may be in reach in the near future. We investigate UHE particle propagation using the Monte Carlo code CRPropa. Particularly, the expected photon fluxes normalized to current experiments as well as prospects for future experiments are illustrated. Varying source and propagation scenarios are analyzed and the impact on secondary GZK photons is shown. For the specific case of Centaurus A, we study which source parameters can be tested by searching for the expected GZK photons.
\end{abstract}

\begin{IEEEkeywords}
 cosmic ray propagation, UHE photon flux, Centaurus A
\end{IEEEkeywords}
 
\section{Introduction}
The origin and nature of the highest energy cosmic rays ($E > 10^{17}$~eV) is still one of the most pressing questions of astroparticle physics. However recent developments show a clear evidence of a suppression in the cosmic ray flux at highest energies. HiRes reported the observation of the GZK cutoff above $\sim 6\cdot 10^{19}$~eV with 5 standard deviation significance \cite{AbbasiHiRes}. Furthermore, the Pierre Auger Observatory rejects the hypothesis that the cosmic ray spectrum continues with a constant slope above $4 \cdot 10^{19}$~eV, with a significance of 6 standard deviations \cite{AbrahamAugerSpec}. 

The composition at these energies still remains a mystery. The Pierre Auger Observatory revealed a correlation between the arrival directions of ultra high energy cosmic rays (UHECR) with energy above $6\cdot 10^{19}$~eV and the positions of active galactic nuclei (AGN) within $\sim 75$~Mpc~\cite{Cronin:2007}. This perhaps indicates a lighter composition since heavier nuclei are more effected by magnetic fields. However, measurements of the depth of shower maximum $X_{\rm max}$ of air showers seem to indicate also a heavier component~\cite{Unger:2007mc}.

In either case, energy loss by propagation effects limit the UHECR horizon\footnote{Here the horizon $d$ is defined as the distance within which 90$\%$ of arriving particles originated.} distance to below $\sim 70$~Mpc at energies $\geq 10^{20}$~eV and give rise to secondary particle production.

\begin{figure}[!t]
\centering
\includegraphics[width=8cm]{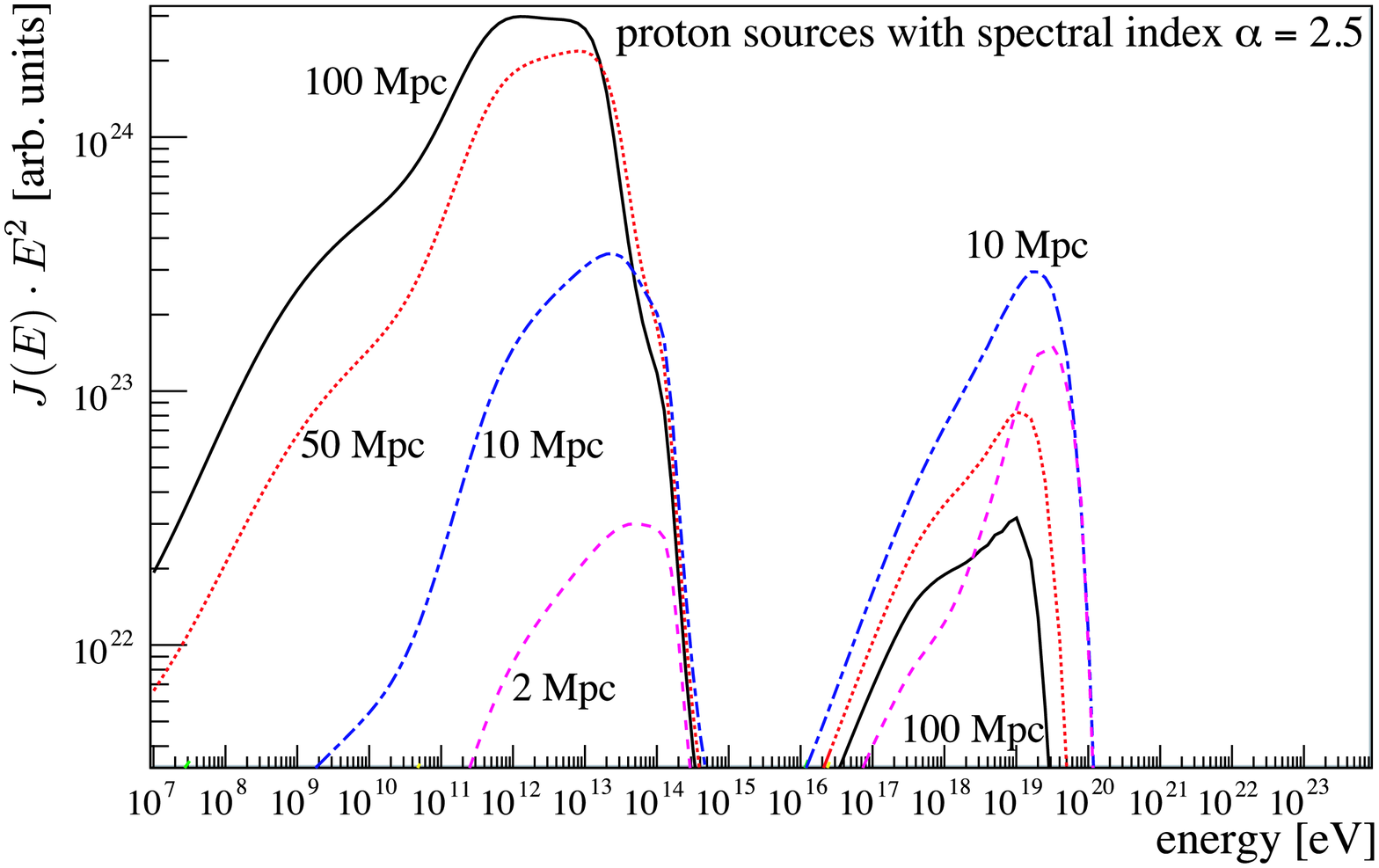}
\caption{Spectrum of secondary photons generated by pion and pair production from a single UHECR \textit{proton} source at a given distance. We consider here a one-dimensional model, with an injection spectral index $\alpha=2.5$ and maximum energy of 10$^{20.5}$~eV. No magnetic fields were taken into account. At a source distance of $\sim 10$~Mpc most of the UHE photons are produced. For closer distances the EM cascade's development has insufficient time to produce a sufficient number of UHE photons whereas for large distances the UHE photon population may cascade down to lower energies (see also \cite{Kuempel2008}).
}
\label{fig_spectrum}
\end{figure}

\begin{figure*}[th]
 \centering
 \includegraphics[width=16.5 cm]{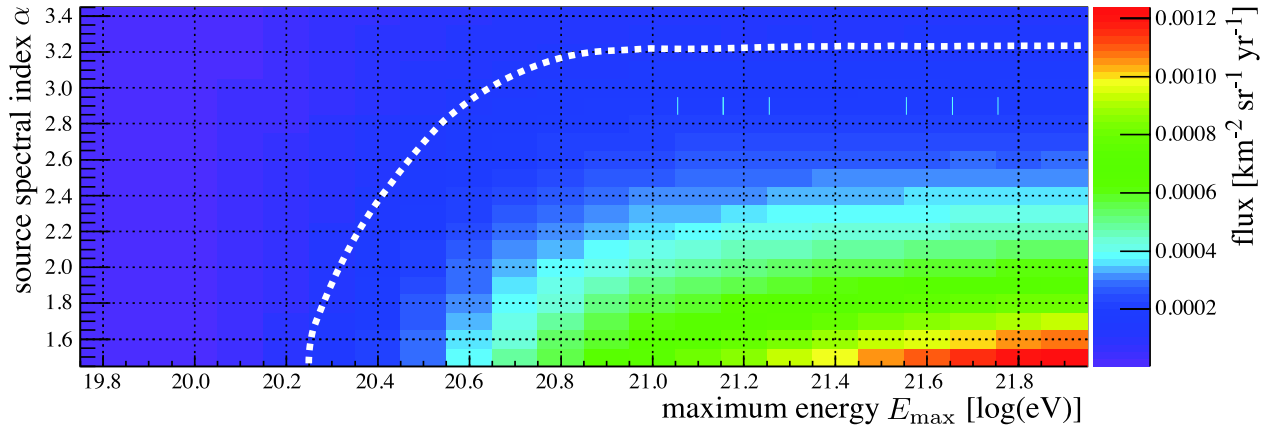}
 \caption{Simulation matrix of varying source parameters. A changing spectral index is shown on the $y$-axis in combination with a varying maximum energy of the source. Color coded is the expected photon flux $F^{10{\rm EeV}}_{\gamma}$ above a threshold energy of 10~EeV. The upper limit on the flux of photons above 10~EeV derived in~\cite{Risse:07} is $F^{\rm limit}_{10{\rm EeV}}=3.8\cdot 10^{-3}$~km$^{-2}$~sr$^{-1}$~yr$^{-1}$. The white dashed line indicates a photon flux level of $F^{\rm limit}_{10{\rm EeV}}/30$, where the factor 30 is a rough estimate of how much the final total exposure of Auger South exceeds the exposure used for $F^{\rm limit}_{10{\rm EeV}}$. One sees that interesting combinations of $\alpha$ and $E_{\rm max}$ can be tested by searching for the expected photon flux.
   }
 \label{fig_flux}
\end{figure*}

To get a clue of an answer of the raised questions it is therefore desirable to expand the knowledge of particle propagation through the local universe. The photon background is a key ingredient for understanding the properties of particle propagation. At energies $\geq 5\cdot 10^{19}$~eV the main channel of energy loss for primary protons is photo-pion production in interactions with background radiation fields which generates the already mentioned GZK feature. Here, the low energy photon can Lorentz transform into a $\gamma$-ray in the rest frame of a very-high energy particle. The cross section increases strongly at the $\Delta^+ (1232)$ resonance. The process can be described as 
\begin{eqnarray*}
 p+\gamma \rightarrow \Delta^{+}(1232) &\rightarrow& n+\pi^+ \\
                                             &\rightarrow& p+\pi^0~.
\end{eqnarray*}

In addition, also further baryon resonances can be excited at increasing energy. The produced neutral pions decay into two UHE photons which in turn are distance limited by $\gamma\gamma$ interactions with background photons. 

In Fig.\ \ref{fig_spectrum} a proton source with spectral index $\alpha=2.5$ was simulated at various distances. The simulations were made using the numerical tool CRPropa~\cite{CRPropa} which is described in more detail in Sec.\ \ref{SecCRPropa}. The resulting UHE photon flux by pion and pair production processes is shown. A region around the source exists where the UHE photon flux is maximal. For closer distances (e.g.\ 2~Mpc) the GZK effect does not yet efficiently produce UHE photons, whereas for larger distances (e.g.\ 50~Mpc) the UHE photon population may go into the development of a full electromagnetic cascade with the main flux arriving at GeV-TeV energies. UHE photons can therefore provide information on local UHECR sources.\\

Up to now no UHE photons were observed yet~\cite{Risse:2007sd}. The strongest constrains on the UHE photon flux at $E > 10$~EeV were set by the Pierre Auger Observatory using surface detector data~\cite{Risse:07}. Recently, new photon limits down to 2~EeV were presented using a combination of surface and fluorescence detector~\cite{Scherini:2009}. These results are statistically limited and can be improved in the future with the capability of measuring the diffuse flux as well as identifying discrete sources. It has also been suggested that a major UHECR flux may arise from just a few nearby AGN such as Centaurus A~\cite{Moskalenko:2008}. At a distance of 3.4~Mpc~\cite{Israel:1998} Centaurus A is by far the nearest active radio Galaxy.

\section{CRPropa simulation}
\label{SecCRPropa}
The interplay between different astroparticle physics experiments has become very important. Existing and planned projects range from UHECR observations like the Pierre Auger Observatory, to neutrino telescopes~\cite{neutrino, neutrino2}, as well as ground and space based $\gamma$-ray detectors operating at TeV and GeV energies, respectively~\cite{gammatel}. Even if a putative source were to produce exclusively UHECR, photo-pion and pair production by protons on the photon background would lead to guaranteed secondary photon and neutrino fluxes that could be detectable. With this motivation a numerical tool called CRPropa~\cite{CRPropa} has been developed that can treat the interface between UHECR, $\gamma$-ray and neutrino astrophysics, and large scale magnetic fields.\\ 

Pion production is modelled by using the event generator SOPHIA~\cite{Sophia} that has been explicitly designed to study this phenomenon and is augmented in CRPropa for interactions with a low energy extra-galactic background light (EBL). Unlike pion production, pair production by protons is taken into account as a continuous energy loss due to the low inelasticity.

The EM cascade code is based on~\cite{Lee96}. All relevant interactions with background photons are taken into account and implemented in CRPropa including single pair production, double pair production, inverse Compton scattering and triplet pair production. There are three different photon backgrounds implemented in CRPropa. The most important one is the cosmic microwave background (CMB) with a well known redshift evolution. For the infrared background a model of Primack \textit{et al.}\ is used~\cite{Primack:2005}. This becomes important for EM cascades around the threshold for pair production and is less significant in the UHE region. Above $\simeq 10^{18}$~eV interactions with the universal radio background (URB) become more important where it can inhibit cascade development due to the resulting small pair production length. We use a model based on observations~\cite{1970Nature}.\\
\begin{figure*}[th]
 \centering
 \includegraphics[width=16.5 cm]{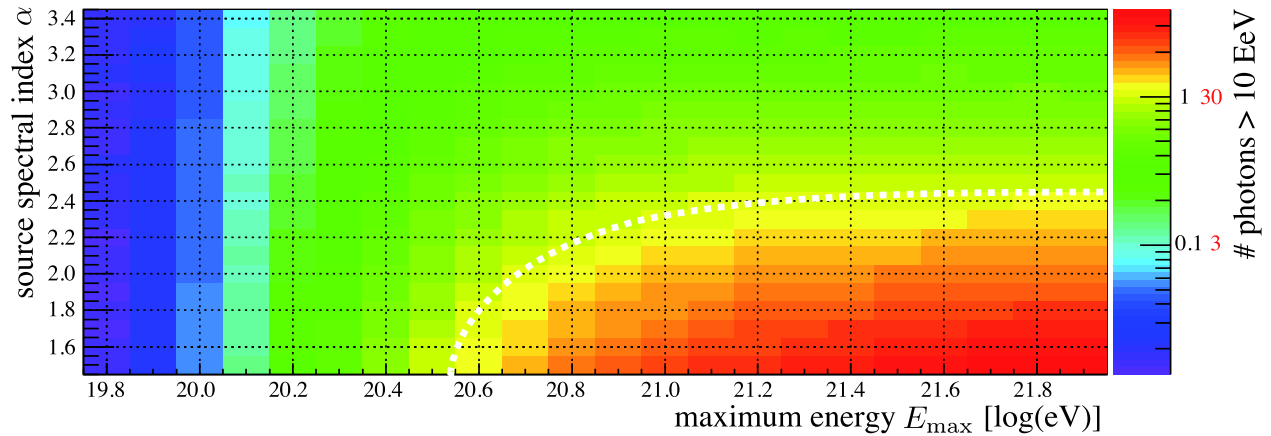}
 \caption{Simulation matrix of varying source parameters. A changing spectral index is shown on the $y$-axis in combination with a varying maximum energy of the source on the $x$-axis. Color coded are the expected number of secondary photons above a threshold energy of 10~EeV for an exposure of 3130~km$^2$~sr~yr, corresponding to~\cite{Risse:07}. The white dashed line indicates one photon above 10~EeV. The red numbers indicate an increased exposure by a factor of 30, corresponding to the expected sensitivity of Auger South.
   }
 \label{fig_NPhotons}
\end{figure*}

In this study we investigate one-dimensional particle propagation for distances of 3.4~Mpc i.e.\ the distance to Centaurus A. Here we assume a proton source with spectral index $\alpha$ and maximum energy $E_{\rm max}$. Protons are injected and propagated (assuming no magnetic field) towards the observer. The resulting EM spectra are recorded and weighted according to the Auger flux spectrum~\cite{Abraham:2008ru} as follows: we assume in a first simplified step that the total flux above 57~EeV originates from a source at a distance of 3.4~Mpc, i.e.\ in this case 27 hadron events above 57~EeV with 7000~km$^2$~sr~yr exposure (as observed in~\cite{Abraham:2008ru}). The ratio $\beta_{7000}$ of the observed 27 events to the simulated total number of arriving particles above 57~EeV is then also used to scale the simulated photon flux above a certain energy $E$ (here we take $E=10$~EeV), i.e.
\begin{equation}
 N^{10{\rm EeV}}_{\gamma,7000} = N^{10{\rm EeV}}_{\gamma,\rm sim} \cdot \beta_{7000}~,
\end{equation}
where $N^{10{\rm EeV}}_{\gamma,\rm sim}$ is the simulated number of photons above 10~EeV and $N^{10{\rm EeV}}_{\gamma,7000}$ the number of photons that are expected to be observed above 10~EeV with an exposure of 7000~km$^2$~sr~yr~\cite{Abraham:2008ru}.\\
The expected integrated $\gamma$-flux above 10~EeV, $F^{10{\rm EeV}}_{\gamma}$, is then calculated via

\begin{equation}
 F^{10{\rm EeV}}_{\gamma} = \frac{N^{10{\rm EeV}}_{\gamma,7000}}{7000~{\rm km^2~sr~yr}}~.
\end{equation}

\section{Results}
In Fig.\ \ref{fig_flux} the expected photon flux $F^{10{\rm EeV}}_{\gamma}$ is shown for varying source parameters $\alpha$ and $E_{\rm max}$. Compared to the upper limit on the flux of photons above 10~EeV derived in~\cite{Risse:07} of $F^{\rm limit}_{10{\rm EeV}} = 3.8\cdot 10^{-3}$~km$^{-2}$~sr$^{-1}$~yr$^{-1}$ all simulated source parameter combinations are compatible with the current upper limit. That is, the present upper limit on the photon flux does not yet constrain Centaurus A as a strong source of UHE protons. We checked that a constant tansverse magnetic field of 100~pG has just a marginal effect on the UHE photon flux in this scenario. The white dashed line in Fig.\ \ref{fig_flux} indicates an improved photon flux level of $F^{\rm limit}_{10{\rm EeV}}/30$, corresponding to a rough estimate of the expected sensitivity of Auger South. Here, certain parameter combinations of $\alpha$ and $E_{\rm max}$ produce a larger photon flux and can thus be tested. 

The number of arriving photons above 10~EeV for varying source parameters that are expected to be observed with an exposure of 3130~km$^2$~sr~yr (corresponding to the exposure used for $F^{\rm limit}_{10{\rm EeV}}$~\cite{Risse:07})
\begin{equation}
 N^{10{\rm EeV}}_{\gamma, 3130} = F^{10{\rm EeV}}_{\gamma} \cdot 3130~{\rm km^2~sr~yr}
\end{equation}
is shown in Fig.\ \ref{fig_NPhotons}. For large $E_{\rm max}$ and small $\alpha$ the expected number increases up to a few photons. For an increased exposure of a factor of 30 the number of arriving photons also scales by a factor of 30 as indicated by the red numbers in Fig.\ \ref{fig_NPhotons}. One sees that, depending on the source parameters, up to several 10 events could be expected.

\begin{figure*}[th]
 \centering
 \includegraphics[width=16.5 cm]{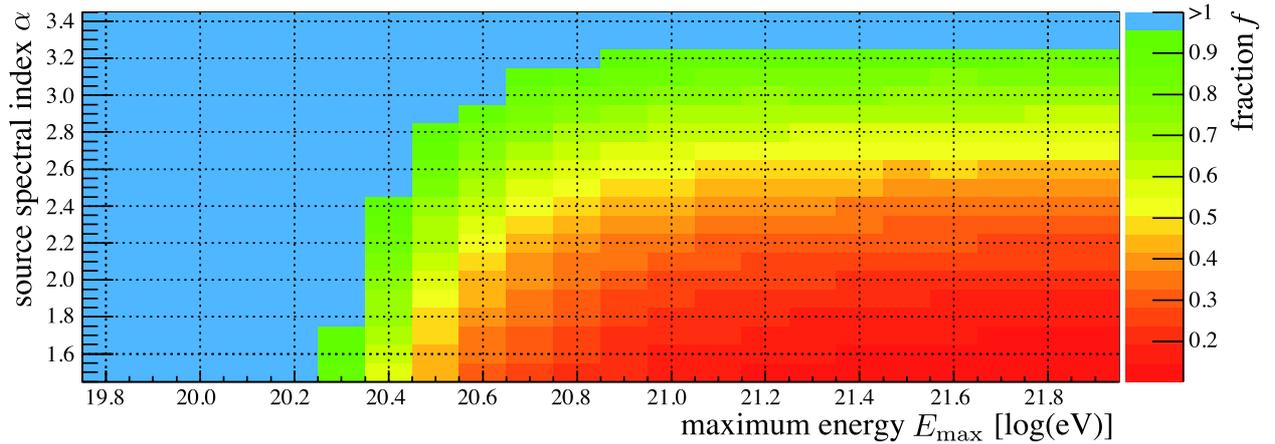}
 \caption{Simulation matrix of varying source parameters. Color coded is the fraction $f=(F^{\rm limit}_{10{\rm EeV}}/30) / F^{10\rm EeV}_{\gamma}(\alpha, E_{\rm max})$ (see also caption of Fig.\ \ref{fig_flux}).
   }
 \label{fig_fraction}
\end{figure*}

So far, we made the simplified assumption that the total flux above 57~EeV is produced by the source $-$ irregardless of the question whether the observed shape of the spectrum or the distribution of arrival directions were reproduced. Relaxing now this assumption one can ask for the fraction
\begin{equation}
f=\frac{F^{\rm limit}_{10{\rm EeV}}/X}{F^{10\rm EeV}_{\gamma}(\alpha, E_{\rm max})}~,
\end{equation}
where $X=1$ refers to the exposure used for $F^{\rm limit}_{10{\rm EeV}}$ and $X \simeq 30$ to the Auger South sensitivity. Assuming now an observed photon upper limit $F^{\rm limit}_{10{\rm EeV}}/X$, for $f > 1$, no constrain on source parameters is possible, while values of $f < 1$ indicate which fraction of the total flux would still be allowed from the source for a given combination of $\alpha$ and $E_{\rm max}$. As is clear from Fig.\ \ref{fig_flux}, $f > 1$ for $X=1$, for all simulated combinations of $\alpha$ and $E_{\rm max}$.

In Fig.\ \ref{fig_fraction}, the case of $X=30$ is shown, and the corresponding fractions of the total flux still allowed by the source can be extracted. For instance, in case of $\alpha \simeq 2$ and $E_{\rm max} \simeq 10^{21}$~eV, no more than $\sim 30\%$ of the total cosmic ray flux could be due to protons from Centaurus A.

\section{Summary}
We studied GZK photon fluxes expected at Earth using the MC program CRPropa. Due to the competition between production of GZK photons (increasing with travel distance of the mother nucleon) and attenuation of GZK photons (increasing with travel distance), the expected photon fluxes show a non-trivial dependence on the source distance (Fig.\ \ref{fig_spectrum}, see also e.g.~\cite{Taylor:2008jz}).

Regarding the specific case of Centaurus A (see also~\cite{Taylor:2009we}), the current photon flux limit~\cite{Risse:07} does not yet constrain Centaurus A as a strong source of UHE protons for the investigated range of spectral indices $\alpha$ and maximum energies $E_{\rm max}$ (Fig.\ \ref{fig_flux}). However, the sensitivity that will be accumulated by Auger South will allow interesting constrains for a broad range of $\alpha$ and $E_{\rm max}$ (Fig.\ \ref{fig_flux} and Fig.\ \ref{fig_fraction}). 

Depending on source parameters, the number of GZK photons above 10~EeV may reach several 10 over the lifetime of Auger South (Fig.\ \ref{fig_NPhotons}). We conclude that the search for UHE photons helps to provide significant clues about the characteristics of potential astrophysical sources.

We note that we regarded only GZK photons, i.e.\ photons produced during the propagation. The photon fluxes may be enhanced in case of interactions {\it at} the source. Such studies, as well as 3d-simulations with CRPropa, are in progress.

\section{Acknowledgment}
We would like to thank G.\ Sigl and E.\ Armengaud for helpful comments and suggestions. This work was partially supported by the German Ministry for Research and Education (Grant 05 A08PX1).

\end{document}